# SUSHI: Sparsity-based Ultrasound Super-resolution Hemodynamic Imaging

Avinoam Bar-Zion, Oren Solomon, Charles Tremblay-Darveau, Dan Adam and Yonina C. Eldar*

*Abstract*—**Identifying and visualizing vasculature within organs and tumors has major implications in managing cardiovascular diseases and cancer. Contrast-enhanced ultrasound scans detect slow-flowing blood, facilitating non-invasive perfusion measurements. However, their limited spatial resolution prevents the depiction of microvascular structures. Recently, super-localization ultrasonography techniques have surpassed this limit. However, they require long acquisition times of several minutes, preventing the detection of hemodynamic changes. We present a fast super-resolution method that exploits sparsity in the underlying vasculature and statistical independence within the measured signals. Similar to super-localization techniques, this approach improves the spatial resolution by up to an order of magnitude compared to standard scans. Unlike super-localization methods, it requires acquisition times of only tens of milliseconds. We demonstrate a temporal resolution of ~25Hz, which may enable functional super-resolution imaging deep within the tissue, surpassing the temporal resolution limitations of current super-resolution methods, e.g. in neural imaging. The sub-second acquisitions make our approach robust to motion artifacts, simplifying *in-vivo* use of super-resolution ultrasound.**

## I. INTRODUCTION

Ultrasound (US) is a cost-effective, reliable, non-invasive and radiation free imaging technique. The use of encapsulated gas microbubbles as contrast agents extends the capabilities of ultrasound to the imaging of fine vessels with low flow velocities. Specifically, contrast-enhanced ultrasound (CEUS) enables real-time hemodynamic and perfusion imaging with high-penetration depth. However, like many non-invasive imaging modalities, spatial resolution limitations prevent classic CEUS imaging from resolving the fine structure of the microvasculature. Therefore, despite their superior penetration depth, classic ultrasound measurements are limited in their capability to detect microvascular changes in response to anticancer [1] and anti-inflammatory treatment [2], and produce functional maps with limited spatial resolution, compared to optical scans [3], [4]. Since anticancer and anti-inflammatory treatments are known to cause structural changes in the microvascular level [5], resolving these vessels could enable fast and direct treatment monitoring capability, valuable for both research and clinical applications. Furthermore, when neural activity is increased, most of the blood flow enhancement is performed in the capillaries [3], so that functional imaging at this level may enable better understanding of neural activity.

Several CEUS super-localization techniques that were introduced in the last few years, were shown to surpass the classic resolution limitations dictated by the point spread function (PSF) of the ultrasound system and provide sub-diffraction scans [6]–[12]. Inspired by optical fluorescent microscopy methods like PALM [13] and STORM [14], these techniques overcome the diffraction limit by capturing a series of frames, each composed of a sparse distribution of microbubbles. The main assumption in these methods is that the microbubbles in each frame are resolvable and therefore the center of each echo is estimated with sub-diffraction resolution. By applying a super-localization procedure and accumulating the localizations over many frames, the overall structure of interest can be revealed. In CEUS super-localization, the assumption of sparse microbubble distribution is satisfied by using very low concentrations of microbubbles [7], [8], [12]; bursting sub-populations of microbubbles [9]; and filtering with spatio-temporal filters to capture sub-sets of microbubbles [6]. By estimating the centers of resolvable PSFs, an improvement of up to 10 fold in spatial resolution was reported [6]. Super-localization has found a range of preclinical applications including 3D anatomical imaging of vessels in the brain [6] and in tumors [10].

Despite recent advances in CEUS super-localization, these methods are still limited by low temporal resolution (typically tens to hundreds of seconds) and large amounts of data (tens of thousands of sequential images) that need to be stored and processed [6], [8]. Reducing these long acquisition times and limiting CEUS super-resolution scans to less than a second is important for two main reasons. First, fast hemodynamic changes, with a timespan of a few seconds and less [15], cannot be captured by current super-localization methods. Therefore, these techniques do not have the temporal resolution needed for functional imaging applications such as functional ultrasound

---

A. Bar-Zion and O. Solomon contributed equally to this work.
This project has received funding from the European Union's Horizon 2020 research and innovation program under grant agreement No. 646804-ERC-COG-BNYQ.
A. Bar-Zion (e-mail: barz@campus.technion.ac.il) and D. Adam (e-mail: dan@biomed.technion.ac.il) are with the Department of Biomedical Engineering, Technion—Israel Institute of Technology, Haifa 32000.

imaging of the brain. Second, long acquisition times make super-localization sensitive to motion artifacts. These artifacts are difficult to compensate for in post processing since the motion is three dimensional and CEUS imaging is commonly performed in 2D [1]. In addition, breath holds have limited efficiency since many patients cannot hold their breaths for long periods of time [16]. Consequently, long acquisition times limit the clinical applicability of CEUS super-localization, especially when imaging internal organs whose scans are highly affected by motion artifacts [17].

Inspired by an optical fluorescence microscopy method named SOFI [18], an alternative approach for improving spatial resolution in CEUS imaging was recently presented [19], termed CEUS SOFI. Relying on the independence of fluctuating CEUS signals in neighboring vessels, high-order statistics of the time-series measured in each pixel are estimated. This technique was shown to produce improvement in spatial resolution that scales with the order of the statistics. Avoiding the assumption of resolvable microbubbles, high ultrasound contrast agents' concentrations were used, enabling short acquisition times and high temporal resolution. However, in practice, the order of the statistics used in SOFI is limited by both the SNR and dynamic range of CEUS signals, restricting the typical resolution improvement to a factor of 2.

Here, we present an approach for fast CEUS super-resolution, termed Sparsity-based Ultrasound Super-resolution Hemodynamic Imaging (SUSHI), which extends the preliminary results presented in [20]. The main goal of SUSHI is fast detection and depiction of hemodynamic changes with sub-diffraction resolution. To this end, SUSHI uses high UCA concentration to maximize the portion of the vasculature imaged during ultrafast acquisitions and makes use of the statistical independence between the fluctuations of CEUS signals originating from different vessels. In addition, SUSHI exploits the sparsity of the underlying vascular structure to improve the spatial resolution beyond the diffraction limit, by relying on sparse recovery techniques [21], [22], [23], [24].

Super-resolution imaging by exploiting sparsity in the correlation domain was recently introduced in the context of fluorescence microscopy, in a method called sparsity-based super resolution correlation microscopy (SPARCOM) [25] [26]. SUSHI extends the ideas of SPARCOM to CEUS imaging by exploiting sparsity within the CEUS correlation domain. However, the signal model in CEUS imaging differs from its optics counterpart, as statistical independence between UCA signals can only be assumed between different blood vessels. Moreover, the CEUS signal includes phase measurements, unlike fluorescent signals which contain magnitude only. SUSHI exploits the phase of the received signal to separate between vessels with opposite flow via Doppler processing, prior to performing sparse recovery. This separation provides additional anatomical structure, based on flow directions, and results in sparser signals for both arteries and veins, compared with the original non-Doppler filtered signal. Sparse recovery in the correlation domain allows SUSHI to reduce acquisition time dramatically and operate with short, sub-second acquisition rates and high UCA concentrations, while achieving spatial resolution close to that of super-localization techniques.

In recent years, sparse representations of signals [27], [28] and the theory of compressed-sensing (CS) [23], [24] have gained popularity and found applications in many research fields such as radar [29], magnetic resonance imaging (MRI) [30], and ultrasound imaging [31]. A discrete signal is said to be sparse, if it can be represented as a linear combination of a small number of functions. That is, in an appropriate transform domain, the signal can be represented by a vector with most coefficients zero, except a small number of non-zero values at unknown locations. CS techniques aim at estimating the locations and values of these coefficients. The theory underlying CS asserts that such a sparse vector can be recovered exactly from a small number of linear measurements taken in a non-adaptive manner [27]. CS has also been used to enable super-resolution in fields such as fluorescence microscopy [25], and coherent diffraction imaging [32], to name a few. In addition, recent works [33], [34] have shown that sparse recovery in the correlation domain may lead to a dramatic increase in the number of detected sources, compared with sparse recovery performed on the signal itself.

The rest of the paper is organized as follows: A parametric model for CEUS signals, including Doppler information, is presented in Section II. Based on this model, statistical independence in CEUS signals is discussed in Section III. SUSHI processing that uses this statistical independence together with sparsity in the underlying vascular structures is described in Section IV. In Sections V and VI, we show improvement in spatial resolution comparable to super-localization, but with much higher temporal resolution. Our results are analyzed and discussed in Section VII. Section VIII concludes the manuscript.

Throughout the paper, $x$ represents a scalar, $\boldsymbol{x}$ a vector and $\boldsymbol{X}$ a matrix. The size of a matrix $\boldsymbol{A}$ is denoted by $MXN$, such that $\boldsymbol{A}$ has $M$ rows and $N$ columns. The $k$th discrete Fourier transform coefficient of $x[p], p = 1, ..., P$ is denoted using capital letters and the index $k$, $X[k], k = 1, ..., P$. The notation $||\cdot||_p$ indicates the $p$-norm. Square brackets $[\cdot]$ relate to discrete-time signals, while round brackets $(\cdot)$ indicate continuous-time signals.

## II. CEUS Signal Model and Doppler Processing

In this section, we specify the analytical model we assume to describe the acquired CEUS signal. This model is then used for Doppler processing, which is a pre-processing step in the overall SUSHI algorithm. It also enables us to exploit the inherent sparsity within the signal to achieve super-resolution, as we describe in Section IV.

### A. Signal model

CEUS imaging is performed by transmitting a series of ultrasound pulses towards the region of interest with uniform time differences $\Delta T$. This series of P measurements is performed over an interval $t \in [0, T]$, where $T = P\Delta T$. The received signals in the transducer elements are focused upon reception in a process called beamforming to produce an ultrasound map of the interrogated tissue and are then demodulated to produce the IQ signal. The IQ signal $f$ is composed of the desired blood signal $b$, resulting from echoes of individual microbubbles, and contaminated by the tissue clutter $c$ and an additive noise component $w$ [35]:

$$f(x, z, t) = c(x, z, t) + b(x, z, t) + w(x, z, t). \quad (1)$$

Here $x$ and $z$ are the lateral and axial coordinates, respectively. Our goal is to utilize the acquired microbubbles signal $b$ to achieve a sub-diffraction representation of the underlying vasculature, with sub-second acquisition times. The first step in our processing scheme is to estimate the relevant blood related component $b$ from $f$. We then exploit the inherent sparse structure of the underlying vasculature, described in Section IV, to achieve fast, super-resolved CEUS imaging.

The removal of clutter noise from CEUS data and the estimation of $b$, denoted as $\hat{b}$, is usually performed based on two priors: the non-linear (harmonic) nature of the echoes produced by the microbubbles at low acoustic pressure, and their distinct velocity patterns. Pulse sequences containing several pulses with different amplitudes and/or phases were developed to separate the non-linear signal component [36]. The echoes resulting from these different pulses are weighted and combined as a preliminary processing step. Next, temporal [37] or spatiotemporal [35] filters can be used to remove the remaining non-linear clutter noise and produce an estimation of the blood related signal $\hat{b}$. IIR filters with projection initialization [37] were used in this work since this method's cutoff parameter is directly connected to the minimal flow velocity in the estimated blood signal and therefore is easier to interpret and control.

We now turn to describe an analytical model for $\hat{b}$, which will enable us to perform additional Doppler processing and to formulate the SUSHI processing. We assume that the ultrasound acquisition is performed by a linear, shift invariant system (LSI). Therefore, at any given time, the filtered blood related signal $\hat{b}$ can be written as a convolution between the reflectivity function of the scanned object, denoted as $i(x, z, t)$, and the point spread function (PSF) of the system $h(x, z)$. Denoting the time-dependent set of detected bubbles at time $t$ as K(t), the reflectivity of the UCAs, $i(x, z, t)$, is modelled as a sum of Dirac delta functions $\delta(\cdot, \cdot)$ at time-varying positions $(x_q(t), z_q(t))$, such that

$$i(x, z, t) = \sum_{q \in K(t)} \delta(x - x_q(t), z - z_q(t)) \sigma_q, \quad (2)$$

where $\sigma_q$ represents the scattering of each bubble. A physical model for the echo from a single bubble can be found in [38]. The signal $\hat{b}$ is the convolution between $i(x, z, t)$, defined in (2), and the PSF, resulting in a stream of pulses model [19]:

$$\hat{b}(x, z, t) = \sum_{q \in K(t)} h(x - x_q(t), z - z_q(t)) \sigma_q. \quad (3)$$

This stream-of-pulses model for CEUS signals is similar to the one in [39].

By discretizing the positions of the microbubbles in (3) and associating them with one of $N_p$ neighboring volume cells in which they are located at a given time, $\hat{b}$ can be described by the following equation [19]:

$$\hat{b}[m\Delta_{xL}, l\Delta_{zL}, t] \approx \sum_{n=1}^{N_p} h[m\Delta_{xL} - x_n, l\Delta_{zL} - z_n] s_n(t), \quad (4)$$

where $m, l \in \{1, \dots, M\}$ and $[\Delta_{xL}, \Delta_{zL}]$ are the indices and dimensions of the pixels in the beamformed image, respectively ($L$ stands for "Low", as in the low-resolution beamformed image); $[x_n, z_n]$ are the positions of the $N_p$ microbubble-containing pixels; and $s_n(t)$ is the time dependent signal, summing the contributions $\sigma_q$ of all the bubbles in each bubble-containing pixel. Here, we consider a square image for convenience only, although the method is easily applicable to rectangular images.

The time dependent fluctuations in each pixel $s_n(t)$ include a multiplicative envelope $a$ and a complex phase whose temporal change over consecutive acquisitions is affected by the velocity of the moving contrast agents. In ultrasound imaging we measure $s_n(t)$ at $t = p\Delta T$, where $p$ is the transmitted pulse index. Following [40] and the derivation in [41] for a single scatterer, $s_n[p\Delta T]$ is described as:

$$s_n[p\Delta T] = a[p\Delta T] \sum_{u \in U_n} \exp^{-j\nu_u p\Delta T + \beta_0}. \quad (5)$$

The set $U_n$ contains the microbubble velocities detected within each volume cell $n$ during the imaging interval, $\nu_u$ are their Doppler angular frequencies, and $\beta_0$ is a random constant phase. Each Doppler angular frequency is related to a specific axial velocity $v_{z,u}$, by the following relation,

$$\nu_u \equiv 2\pi(2f_0 v_{z,u} / c). \quad (6)$$

Here, $c$ is the velocity of sound in the medium and $f_0$ is the center frequency of the ultrasound wave. A single volume cell $n$ can contain bubbles belonging to several vessels or streamlines with different Doppler angular frequencies $\nu_u$. In previous works, the random envelop $a$ was shown to follow the K-distribution [42].

The complex phase in (5) enables the use of Doppler processing, not only to remove clutter noise but also to decompose the CEUS signal according to the blood flow velocities within the vessels. By doing so, CEUS signals that belong to different blood vessels with distinct Doppler signals can be separated even if they spatially overlap. This Doppler processing is performed on each pixel's time-trace separately.

### B. Doppler processing

Next, we detail our Doppler pre-processing, starting with a single pixel, whose time trace is given by (4), sampled at times $t = p\Delta T$. In the following, the $\Delta T$ notation is removed, for convenience. Considering (4) at discrete time-points $p$, we obtain the $P$-point temporal discrete Fourier Transform (DFT) of $\hat{b}$ in pixel $[m, l]$ as

$$B_{m,l}[k] = DFT\left\{ \sum_{n=1}^{N_p} h[m\Delta_{xL} - x_n, l\Delta_{zL} - z_n] s_n[p] \right\}$$
$$= \sum_{n=1}^{N_p} \sum_{u \in U_n} h[m\Delta_{xL} - x_n, l\Delta_{zL} - z_n] A_u[k], \quad (7)$$

where the DFT of the $u$th velocity component from (5) $(DFT\{s_n[p]\})$ is given by

$$A_u[k] = DFT\left\{a[p]\, e^{-j\nu_u p \Delta T + \beta_u}\right\}[k], \quad k = 0, ..., P-1. \quad (8)$$

The double sum in (7) presents the contribution of the velocity distribution $U_n$ in each bubble-containing location $n$. As part of our Doppler processing we divide the continuous Doppler spectrum of the imaged blood vessels into $U$ bands. By doing so and looking at the entire scan instead of a single pixel, we can reorder the double sum in (7) according to a desired number of bands $U$, and the number of volume cells containing microbubbles in each Doppler band $N_u$. Subsequently, the signal $B_{m,l}[k]$ is decomposed according to its spectral content:

$$B_{m,l}[k] = \sum_{u=1}^{U}\left(A_u[k]\sum_{n \in N_u}h[m\Delta_{xL} - x_n, l\Delta_{zL} - z_n]\right). \quad (9)$$

Since we divide $B_{m,l}[k]$ into $U$ different Doppler bands (which together cover the entire spectrum), we may apply a series of temporal bandpass filters to the time series in each pixel, to separate the bands. By doing so, the blood signal $\hat{b}$ at pixel $[m, l]$ is decomposed into $D_f$ signals, each denoted by $\hat{b}^d$, with different flow characteristics:

$$\hat{b}_{m,l}^{d}[p] = DFT^{-1}\left\{B_{m,l}[k]\,\Pi_{m,l}^{d}[k]\right\}[p], \quad (10)$$

where $\Pi^d$, $d = 1, ..., D_f$ correspond to the series of $D_f$ temporal bandpass filters that together cover the relevant Doppler frequency bands. These temporal filters are applied to each pixel's time-trace, in parallel. When applied to all the pixels in the scan, this Doppler processing results in $D_f$ movies, each showing different flow patterns and sparser vasculature, divided according to the different Doppler velocities. In this work, we chose $D_f = 2$, and used one filter to separate the positive frequencies and another to separate negative frequencies, with respect to the transducer. Thus, arterial and venous vasculature could be separated (as presented in Fig. 2), according to the direction of flow. An illustration of the temporal filters is given in Appendix A1, Fig. 5.

In the time domain, the filtered signal $\hat{b}^d$ in (10) is given by

$$\hat{b}^d\left[m\Delta_{xL}, l\Delta_{zL}, p\right] = \sum_{n=1}^{N}h[m\Delta_{xL} - x_n, l\Delta_{zL} - z_n]\,s_n^d[p], \quad (11)$$

with $s_n^d[p]$ being the time dependent signal fluctuation in each bubble containing pixel $n$, and $d$ being the index of the Doppler band. From here on, the processing is performed on each filtered signal, $\hat{b}^d$ and we omit the superscript $[\cdot]^d$ for ease of notation. In the next section we describe how to exploit statistical properties of CEUS fluctuations over each filtered signal to improve the spatial resolution.

## III. STATISTICAL PROCESSING OF CEUS TIME-SERIES

The statistical blinking of CEUS signals was recently utilized in a method called CEUS SOFI [19], to enhance the spatial resolution of these scans while maintaining high temporal resolution. In CEUS SOFI, the moments of the time-series in each pixel were presented together as 2D images. SUSHI improves upon CEUS SOFI, by exploiting sparsity in the correlation domain on each filtered signal. In this section we briefly describe the main ideas behind the statistical processing of CEUS scans as presented in [19]. In Section IV inspired by the SPARCOM method developed for super resolution florescent microscopy [25] we exploit sparsity in the correlation domain to improve the resolution even further. This enables to achieve super-resolution imaging while dramatically reducing the acquisition time.

We make the following assumptions throughout:

1. For each processed ensemble, the statistics of the measured echoes from the bubbles in each volume cell $s_n^d[p]$ depend only on the time difference $\tau$ between the measurements and not on the specific measurement times (i.e. it is a wide sense stationary process).

2. The location of a vessel containing volume cells positioned around $[x_n, z_n]$, $n = 1, ..., N_P$ does not change during the short acquisition time of the processed ensemble (and so, no image registration is required).

3. Temporal signal fluctuations in volume cells that belong to different blood vessels are statistically independent.

Following assumptions 1-3, SOFI processing calculates statistical quantities of the CEUS time-series in each pixel and presents them as a single image. For example, the 2nd order SOFI signal $g_2$ is produced by estimating the auto-correlation of each pixel in (11) for a prechosen discrete time-lag $\tau$ [19]:

$$g_2\left[m\Delta_{xL}, l\Delta_{zL}, \tau\right] = \sum_n\left|h[m\Delta_{xL} - x_n, l\Delta_{zL} - z_n]\right|^2 g_n[\tau]$$

$$+ \sum_{\substack{i,l \\ i \neq l}}h[m\Delta_{xL} - x_i, l\Delta_{zL} - z_i]\cdot \bar{h}\left[m\Delta_{xL} - x_l, l\Delta_{zL} - z_l\right]g_{il}[\tau], \quad (12)$$

where $g_n[\tau] = E\{\tilde{s}_n[p + \tau]\overline{\tilde{s}_n[p]}\}$ is the autocorrelation function of the temporal fluctuations of pixel $n$, and $g_{il}[\tau] = E\{\tilde{s}_i[p + \tau]\overline{\tilde{s}_l[p]}\}$ is the cross-correlation of pixels $i$ and $j$. Here $\tilde{s}_n[p] = s_n[p] - E\{s_n[p]\}$, $n$ is the pixel index, $\tau$ is the discrete pre-determined delay of the autocorrelation function, and $i$ and $l$ are indices of (dependent) volume cells located in the same streamline. Bubbles flowing independently in different vessels produce only squared absolute-valued PSF, seen in the first expression in (12) but not cross-terms, seen in the second part of this equation. Being narrower than the original PSF, the squared absolute-valued PSF represents the improved separation between the vessels. The second term shows the wider first order PSFs that smooth the signal from microbubbles flowing along the same vessels. In [19] higher statistical orders were used to further increase the resolution of SOFI images.

The application of SOFI to CEUS scans provides several advantages including a significant SNR improvement [43] and increase in spatial resolution which scales as the square root of the applied statistical order [19]. However, in practice, high order statistics beyond the 4th moment are rarely used since longer ensembles are needed for estimating high order statistics; in addition, when using high order statistics strong echoes mask weaker echoes from adjacent bubbles [18]. The limited improvement in spatial resolution achieved by CEUS SOFI and the ideas presented in SPARCOM [25] motivate the combination of the statistical priors of SOFI with additional priors on the characteristics of the underlying signal of interest, leading to the SUSHI framework for super-resolution CEUS imaging. Using this approach, the vasculature is reconstructed on a grid denser than the grids of CEUS SOFI images, without increasing the required scan time (e.g. Fig. 1).

## IV. SUSHI PROCESSING

### A. Exploiting sparsity for super-resolution

We now describe the SUSHI processing in detail which exploits sparsity of the underlying vasculature in the correlation domain. Sparse recovery processing is demonstrated by using the correlation-based images calculated from the low-resolution measurements. The key idea in SUSHI is to model the underlying vasculature as composed of point-targets on a higher resolution grid. We assume that on this grid, the underlying vasculature is sparse. This assumption leads to the formulation of a sparse recovery problem, which is solved with a numerically efficient algorithm. SUSHI can easily be expanded to images of higher-order statistics as well (see discussion in Appendix A2).

Consider the correlation based CEUS-SOFI image (12). The first term represents the auto-correlation of the temporal fluctuations arising from each microbubble, while the second term constitutes the cross-correlation function of temporal fluctuations from adjacent microbubbles flowing within the same vessel in a correlated way. The second term was neglected in the following processing, as it is does not affect the support of the vessels. By posing a sparse recovery problem on the correlation image, we recover a super-resolved map of the vasculature from scans with overlapping CEUS echoes.

To achieve super-resolution, we introduce a new high-resolution grid with spacing $[\Delta_{xH}, \Delta_{zH}]$, such that $[x_n, z_n] = [i_x \Delta_{xH}, i_z \Delta_{zH}]$ for some $i_x, i_z \in \{0, \ldots, N-1\}$, while $[\Delta_{xL}, \Delta_{zL}]$ is referred to as the low-resolution grid. We assume that $\Delta_{xL} = D \, \Delta_{xH}$ and $\Delta_{zL} = D \, \Delta_{zH}$ for some $D \geq 1$, and consequently it holds that $N = DM$ (in all of our experiments we fix $D = 8$). Thus, we start from an $MXM$ CEUS correlation image and reconstruct an $NXN$ super-resolved image which is $D^2$ times larger. When studying the structure of the vasculature residing in the imaged plane, the underlying information we wish to obtain is the set of voxels which contain vessels in this high-resolution grid.

Omitting the cross-correlation term from (12) (applied to each $\hat{b}^d$ ), we rewrite it in Cartesian form as

$$g_2\big[m\Delta_L, l\Delta_L, \tau\big]$$
$$= \sum_{i_x, i_z} \big| h\big[m\Delta_L - i_x\Delta_H, l\Delta_L - i_z\Delta_H\big]\big|^2 \, g_{i_x, i_z}[\tau]. \quad (13)$$

Substituting $\Delta_{xL} = D \, \Delta_{xH}$, $\Delta_{zL} = D \, \Delta_{zH}$ into (13) we have

$$g_2\big[mD, lD, \tau\big] = \sum_{i_x, i_z = 0}^{N-1} \big| h\big[mD - i_x, lD - i_z\big]\big|^2 \, g_{i_x, i_z}[\tau], \quad (14)$$

where $\Delta_{xL}, \Delta_{zL}, \Delta_{xH}$ and $\Delta_{zH}$ are omitted for convenience.

Since we discretized the possible positions of vessels within $NXN$ pixels, $g_{i_x, i_z}[\tau]$ represents the autocorrelation of pixel $[i_x, i_z]$ on the high-resolution grid, for a prechosen time-lag $\tau$. The possible locations of the vessels are discretized according to the high-resolution grid $i_x, i_z \in [0, \ldots, N-1]$, such that if no bubble is present in some pixel, then its autocorrelation will be zero. By estimating the locations in which $g_{i_x, i_z}[\tau] \neq 0$, a high-resolution estimation of the vascular structure can be achieved. For example, choosing $\tau = 0$, the variance of the fluctuations is estimated using the high resolution grid. Each pixel in the recovered image corresponds to the variance of the echoes originating from this point (or zero, if no echoes are detected).

The above model and the sparsity prior on the underlying vasculature, enable us to estimate the locations of the vessels by solving an inverse problem as described below. Following a similar line of computation to that presented in [25] we consider (14) in the discrete Fourier domain, leading to an efficient numerical estimation of the high-resolution image. Note that $g_2[mD, lD, \tau]$ is an $MXM$ matrix. We denote its 2D discrete Fourier transform by $G_2[k_m, k_l, \tau]$, where $k_m, k_l$ are $MXM$ spatial frequencies. Performing an $MXM$ 2D DFT on (14) yields

$$G_2\big[k_m, k_l, \tau\big] = H\big[k_m, k_l\big] \sum_{i_x, i_z = 0}^{N-1} g_{i_x, i_z}[\tau] e^{-j\frac{2\pi}{N} k_m i_x} e^{-j\frac{2\pi}{N} k_l i_z}, \quad (15)$$

where $H[k_m, k_l]$ is the $MXM$ 2D DFT of the $MXM$ squared, absolute value PSF $|h(xD, yD)|^2$.

Next, we rewrite (15) in matrix-vector notation. To simplify the equation, we perform column-wise stacking (vectorization) of $G_2[k_m, k_l, \tau]$, and denote the result as an $M^2$ long vector $\boldsymbol{y}[\tau]$, that is $\boldsymbol{y}[\tau] = vec\{G_2[k_m, k_l, \tau]\}$. In a similar manner, we vectorize the $NXN$ image statistics on the high-resolution grid $g_{i_x, i_z}[\tau]$ and denote the result as an $N^2$ long vector $\boldsymbol{x}[\tau]$, so that $\boldsymbol{x}[\tau] = vec\{g_{i_x, i_y}[\tau]\}$, $i_x, i_z = 0, \ldots N-1$. Thus, $\boldsymbol{x}[\tau]$ represents the underlying vasculature we wish to recover on the high-resolution grid, and is assumed to be sparse. Rewriting (15) in matrix-vector form yields:

$$\boldsymbol{y}[\tau] = \boldsymbol{H}\left(\boldsymbol{F}_M \otimes \boldsymbol{F}_M\right) \boldsymbol{x}[\tau] = \boldsymbol{A}\boldsymbol{x}[\tau], \qquad \boldsymbol{A} \in \mathbb{C}^{M^2 X N^2}. \quad (16)$$

Here, $\boldsymbol{A} = \mathbf{H}(\boldsymbol{F}_M \otimes \boldsymbol{F}_M)$, $\boldsymbol{H}$ is an $M^2 X M^2$ diagonal matrix with diagonal elements $\{H[0,0], \ldots, H[M-1, M-1]\}$, $\otimes$ stands for the Kronecker product and $\boldsymbol{F}_M$ denotes a partial $MXN$ DFT matrix, created by taking the rows corresponding to the lowest $M$ frequency components of a full $NXN$ DFT matrix. Additional illustrations and information regarding the Doppler processing and the construction of $\boldsymbol{F}_M$ (Fig. 6) and $\boldsymbol{A}$ are presented in Appendixes A1 and A2.

We solve for $\boldsymbol{x}[\tau]$ in (16) by considering the following optimization problem which includes the prior that $\boldsymbol{x}$ is sparse:

$$\min_{\boldsymbol{x}} \lambda \parallel \boldsymbol{x} \parallel_1 + \frac{1}{2} \parallel \boldsymbol{y} - \boldsymbol{A}\boldsymbol{x} \parallel_2^2. \quad (17)$$

Here $\tau$ is omitted for the sake of simplicity and $\lambda \geq 0$ is a regularization parameter. Exploiting sparsity enables reconstruction of the underlying vascular structure at sub-diffraction resolution even without separation of single bubbles. If $\tau = 0$, then $\boldsymbol{x}$ represents the variance of the CEUS signal fluctuations, which is a non-negative quantity, and consequently the constraint $\boldsymbol{x} \geq 0$ is added.

Many existing algorithms aim at solving (17). We focus on the FISTA [21] algorithm which is known to achieve the fastest possible (worst-case) convergence rate for a first-order method, as described by Nesterov [21]. Since we formulated our problem in the (discrete) frequency domain, we are able to evaluate the application of $A$ on a vector using fast Fourier transform (FFT) operations. A detailed description of the SUSHI algorithm is given in Algorithm 1. The $sign$ function in line 2 of the algorithm operates elementwise and is equal 1 for a positive input, zero for a zero input and $-1$ for a negative input. The calculation of the gradient in Algorithm 1, which involves the application of $A^T$ and $A^T A$ on vectors, can be computed very efficiently, due to our formulation of (16) in the discrete frequency domain (the explicit expression of the gradient is given in step 1 of Algorithm 1). In particular, as shown in detail in [26] in the context of SPARCOM, the matrix $A$ does not need to be stored in memory; rather it can be applied directly using fast Fourier transform (FFT) operations. Similarly, in the Fourier domain, $A^T A$ admits a structure known as block circulant with circulant blocks (BCCB), so that it can be applied on vectors directly again using only FFTs leading to a very efficient numerical implementation.

Sparse recovery and in particular super-resolution recovery can also be performed by assuming that $x$ is sparse under different transformations. For example, sparse representation of vascular structures in Haar and Daubechies wavelets has been used in MRI compressed sensing algorithms [30]. Another option is to consider super-resolution recovery under total-variation [44] (TV) constraints,

$$\min_x \lambda \text{TV}(x)_1 + \frac{1}{2} \| y - Ax \|_2^2 . \quad (18)$$

In its general form, we utilize an analysis based formulation

$$\min_x \lambda \| T^* x \|_1 + \frac{1}{2} \| y - Ax \|_2^2, \quad (19)$$

where $T$ stands for the desired transformation such as the discrete wavelet transform or DCT and $(\cdot)^*$ stands for the adjoint operation. Specifically, (19) implies that $x$ has a sparse representation under the transformation $T$, that is $x = T\alpha$, and $\alpha$ is sparse. In the case of a TV prior we used the TV-FISTA

---

**Algorithm 1:** SUSHI via FISTA for minimizing (17)

**Input:** CEUS movie $\hat{b}$ with $T$ frames, regularization $\lambda > 0$, maximum iterations number $K_{MAX}$, time-lag $\tau$

**Output:** $x_{K_{MAX}}$

**Statistical pre-processing:**

For each pixel in $\hat{b}$ estimate its correlation:

$$g_2[m, l, \tau] =$$
$$\frac{1}{T-\tau} \sum_{t=1}^{T-\tau} \hat{b}[m, l, t] \overline{\hat{b}[m, l, t+\tau]} - \frac{1}{T^2} \left( \sum_{t=1}^{T} \hat{b}[m, l, t] \right)^2$$

Vectorize $g_2$: $y = vec\{g_2\}$

**Initialize:** $z_1 = x_0 = 0$, $t_1 = 1$, $k = 1$ and $L_f = \|A^T A\|_2$

**While** $k \le K_{MAX}$ or stopping criteria not fulfilled

1. Calculate $q_k = A^T A z_k - A^T y$ as described in [26]

2. $x_k = \max \left( \left| z_k - \frac{1}{L_f} q_k \right| - \frac{\lambda}{L_f}, 0 \right) \cdot \text{sign}(z_k - \frac{1}{L_f} q_k)$

3. $t_{k+1} = 0.5(1 + \sqrt{1 + 4t_k^2})$

4. $z_{k+1} = x_k + \frac{t_k - 1}{t_{k+1}}(x_k - x_{k-1})$

5. $k \leftarrow k + 1$

**End**

---

[45] formulation to solve the minimization problem, while for the analysis problem we used the S-FISTA [22] algorithm. The SUSHI images in this paper are generated by solving (17). Comparison to (18) and (19) can be found in Appendix A3. Solutions of (17) and (19) showed similar reconstruction results, which were better in our examples than piece-wise-constant images obtained by minimizing (18).

### B. PSF Estimation

In practice, to apply the matrix $A$ in (17), the PSF $|h|$ must be estimated first. Generally, even when using high concentrations of ultrasound contrast agents, echoes from resolvable bubbles can be expected, at least in small blood vessels with low density flow. In this work, these resolvable echoes were exploited for PSF estimation, using a three-step process. First, the correlation between each image patch and an $MXM$ template patch is calculated. The template patch can be either manually picked or computed according to the geometry of the transducer and the imaging depth. Patches whose correlation with the template patch are above a predefined threshold are considered relevant. These $L$ patches are automatically aligned to the template using rigid body registration and stacked together to produce an $MXMXL$ matrix. Finally, the $MXM$ PSF is estimated by taking the mean of each pixel, over the $L$ patches. As initialization, patches containing resolvable microbubbles were selected from the *in-vivo* scans. The mean of these patches was used as a template for further automatic patch selection.

## V. MATERIALS AND METHODS

We validate SUSHI using numerical phantoms and *in vivo* scans. In these tests, SUSHI achieves spatial resolution gain comparable to that of super-localization approaches, but with short acquisition intervals of only $40-60ms$, compared to tens or hundreds of seconds in super-localization [6], [8]. Using SUSHI, we demonstrate for the first-time sub-second hemodynamic changes in a vascular scan of a rabbit kidney with sub-diffraction spatial resolution. Selected frames from a time-lapse movie with a frame-rate of $25Hz$, similar to the frame-rate of clinical hemodynamic Doppler scans [4], are presented in Section VI.

### A. Simulations

We first test SUSHI using a numerical phantom simulation, to investigate its possible resolution gain and stability under different concentration of microbubbles. To compare the results of the proposed method to a known underlying geometry, a 2D 150 frame long numerical simulation of signals originating from two close by parallel vessels is performed. Three Gaussian bubbles were positioned in the right vessel and six in the left vessel. Additional simulations with up to 4-times the number of bubbles tested the effects of higher concentrations (see also additional concentrations in Fig. 10). Here, Doppler processing was not applied, in order to test the separation of vessels with similar flow velocities using sparsity alone. The dimensions of the simulated Gaussian PSF were set according to the experimental PSF estimated from a rabbit kidney scan, as presented in Appendix A2. The standard deviations of these microbubbles is defined as $\sigma_x = 0.27 \ mm$ and $\sigma_z = 0.23 \ mm$ in the lateral and axial directions, respectively. The centers of the simulated vessels are separated by $1.2 \cdot \sigma_x$ to make sure they are non-resolvable in the temporal mean image. The velocity in the left vessel is defined as $0.25\sigma_z/frame$ while the velocity in the right vessel is $0.5\sigma_z/frame$. In this scenario, due to the velocity differences, in some frames microbubble overlap is present, such that two microbubbles are horizontally aligned, and their centroids distance is $1.2 \cdot \sigma_x$. Initial distance between the bubbles was chosen randomly with a uniform distribution, while preventing the event of connected bubbles. The autocorrelation of the simulated complex signals is calculated together with the temporal mean of the envelope signal in each pixel.

SUSHI images are also compared to super-localization images generated throughout the paper using the Image-J software [46] and the ThunderSTORM plug-in [47]. In essence, this plug-in finds local maxima points and performs a non-linear fit to a Gaussian for each such detection, to achieve sub-pixel precision. In an iterative manner the Gaussian width is also estimated in the process. Specifically, in this study we performed this fit via the weighted least squares option. Prior to the fitting procedure, this code denoised the input data using a wavelet based built-in denoising procedure with B-spline of order 2 and scale 3. Both SUSHI and the super-localization recovery were smoothed, as is customary in single molecule microscopy, with the same Gaussian kernel. A bifurcation simulation similar to the first bifurcation in [48] was also performed (see Fig. 11).

### B. In-vivo Ultrafast Imaging

The *in vivo* scans presented throughout the paper belong to a New Zealand white rabbit model: plane wave pulse inversion Doppler (PID) [39] was used to image normal vessels in the kidney of healthy rabbits. All the scans were acquired using a clinical Aixplorer ultrasound system (Supersonic Imagine, Aix-en-Provence, France) and an L15-4 linear probe transmitting at a PRF of 5KHz. Only the central part of the elements is used upon reception due to channel limitation. The carrier frequency of the transmitted single cycle pulses was 4.5MHz and a mechanical index (MI) of 0.06 was used to reduce the burst rate of the microbubbles. Definity (Lantheus Medical Imaging Inc., N. Billerica, MA, USA), a clinically approved contrast agent, was used at a clinical concentration of 10 µL/Kg. Following the injection of 0.5mL of this contrast agent into the ear vain of the rabbit, an additional 1mL of saline was used to flush it. All the protocols were approved by the Sunnybrook Research Institutional Review Board.

The initial step in the processing of the received IQ data was the weighted summation of the echoes from the different pulses that compose the PID sequence. This process cancels out the linear part of the signal and maintains the non-linear bubble related signal. The single channel data is then beamformed to produce a 2D + time cine matrix. Fast time (axial direction) 2-12MHz FIR band-pass filter was applied to remove noise outside the pass band of the transducer. Next, Doppler (slow time) processing was performed to remove the static clutter from the moving blood and to provide separation between small vessels with blood flowing in opposite directions. A sixth order Butterworth filter with a stopband of 0.03PRF was used as the clutter filter (wall filter). This threshold is selected since it provides good separation between flow and clutter.

## VI. RESULTS

### A. Simulations

The numerical simulation covered a wide range of contrast agents' concentrations. In practice, even when high concentrations of contrast agents are injected and many microbubbles are found in the imaging plane, separated microbubbles can be frequently found in specific locations containing small vessels. The simulation depicted in Fig. 1 a-d, represents a CEUS patch with such low concentration of microbubbles within two adjacent vessels. Figure 1a shows the temporal mean image, created by averaging the envelope of all 150 frames included in the movie. Figure 1b illustrates $2^{nd}$ order SOFI reconstruction (zero time-lag), and Figs. 1c and 1d show the SUSHI and super-localization reconstructions, respectively. In contrast to Figs. 1a-d, Figs. 1f and 1g depict the SUSHI and super-localization reconstructions for high local concentration, respectively. Judging the panels, SOFI resulted only in a slight decrease in the intensity between the two streamlines, while SUSHI and the super-localization technique fully resolved them. Figure 1e provides further support to this conclusion, by presenting an intensity cross-section along the lateral direction for the low concentration scenario (yellow line in Figs. 1a-1d). The plus-head (black line) in Fig. 1e corresponds to the temporal mean image, diamond-head (blue line) to $2^{nd}$ order SOFI image and the right-head (green line) and left-head (turquoise line) to the SUSHI and super-localization images, respectively. Both SUSHI and super-localization techniques resolve the two lines, while the $2^{nd}$ order SOFI does not (a very minor dip is present). Furthermore, the SUSHI and super-localization profiles depict an almost perfect match.

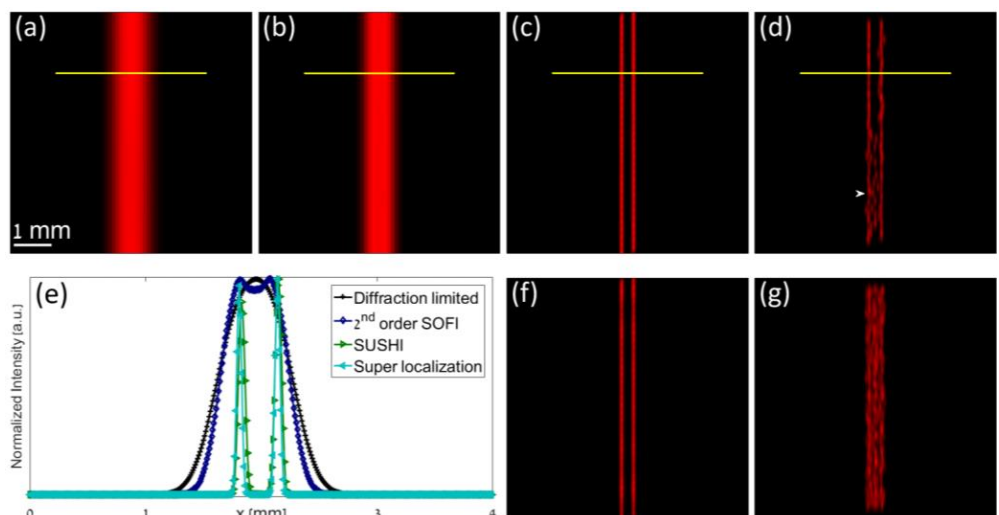

Fig. 1: **Simulation showing the resolution capabilities of SUSHI.** Flow of low concentration of microbubbles within two nearby narrow tubes is depicted by temporal mean (150 frames) (a), $2^{nd}$ order SOFI (b), SUSHI and super-localization ((c) and (d), respectively) compared to higher concentration SUSHI and super-localization ((f) and (g), respectively) reconstructions. Panel (e) illustrates the intensity profiles measured along the horizontal yellow lines. The measured ratio between the temporal mean FWHM to the SUSHI/super-localization FWHM is 9.96 and 12.59, respectively, implying an order of magnitude improvement over that of the temporal mean image. Finally, the SUSHI recovery appears smoother compared to the super-localization recovery for both low and high concentrations, without the false-positive detections appearing in super-localization, even at relatively low microbubble concentration (and more so in high concentrations).

To quantify the resolution enhancement, the full width at half maximum (FWHM) ratio between that of the temporal mean and SUSHI/super-localization profiles (right peak) is measured as 9.96 and 12.59 respectively. The SUSHI FWHM ratio implies a resolution increase by an order of magnitude beyond the classical resolution of the scanner, similar to the results presented previously using super-localization [8].

Nevertheless, it is important to note that even at this simulation of low contrast agent concentration, super-localization processing produces false positive detection, marked by the white arrow at the lower part of Fig. 1d, which are absent in the SUSHI image. Similar results were also observed in the bifurcation simulation (Appendix A4, Fig. 11). This effect worsens as the concentration increases (panels 1g and 1f). The rate of false positive detections increases with the bubbles concentration for super-localization (panel 1g), while the SUSHI recovery (panel 1f) seems similar to the recovery for low concentrations. We conclude that SUSHI operates better in higher concentrations than super-localization. Thus, SUSHI is able to process scans of higher concentrations of contrast agents, which leads to a dramatic reduction in the acquisition time, while producing super-resolved images of the vasculature.

### B.  In-vivo Experiments

To validate the performance of SUSHI when applied to *in vivo* scans, two scans previously presented [19], [49] are reprocessed. These scans include noise, out-of-plane reflectance and depict the true non-linear behavior of the injected microbubbles at high ultrasound contrast agent concentrations usually used in the clinic.

The capability of Doppler processing to separate between adjacent vessels with different flow velocities is demonstrated in Fig. 2. Here, SUSHI is implemented on a rabbit kidney scan containing 150 frames. The temporal mean image of the kidney is presented for reference in Fig. 2a. The $2^{nd}$ order correlation images of the positive and negative flow components are displayed in Fig. 2c and Fig. 2d, respectively. These two images show different vascular structures with higher resolution compared to the temporal mean image. Figure 2b presents the SUSHI reconstruction of the entire kidney, divided into both flow directions. This panel serves to emphasize the power of SUSHI to process entire organs, without any restrictive assumptions on the geometry of the blood vessels. By examining Figs. 2a-2d, it can be seen that indeed, the intertwined vasculature of arteries and veins is separated. Therefore, Doppler based filtering serves as an important pre-processing step. CS processing is then performed on each image separately, to achieve better depiction of the blood vessels, and their flow directions. Although some of the largest blood vessels appear slightly grainy, we emphasize that these blood vessels are considerably larger than the diffraction limit, and are not the focus of the resolution enhancement performed by SUSHI. Improving their visual quality is a matter for future work. To gain a detailed comparison between the spatial resolution enhancement of the different approaches, we turn to perform our analysis on specific regions of interest.

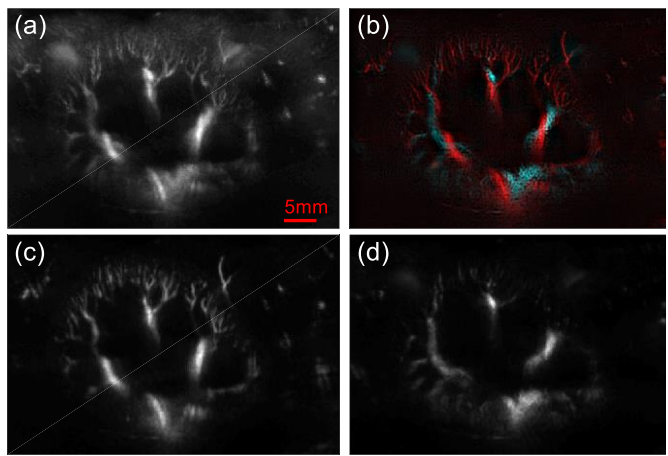

Fig. 2: **Decomposition of the vasculature using Doppler pre-processing.** (a) Temporal mean image of the kidney. (b) SUSHI reconstruction. Red indicates negative flow and cyan positive flow with respect to the transducer. (c) Correlations SOFI image of the negative flow. (d) Correlations SOFI image of the positive flow. Panels (c) and (d) display nearby vessels with opposite flow velocities are decomposed using Doppler processing. Comparing the panels, Doppler processing can clearly distinguish between overlapping blood vessels according to their flow directions. Moreover, panel (b) shows clear SUSHI high-resolution reconstruction of the entire kidney with separation to positive and negative flows.

Figure 3 presents a comparative processing of a patch taken from the kidney scan presented in Fig. 2. In this scan, UCA overlap is present in most frames. Panels (a)-(d) show temporal mean, $4^{th}$ order SOFI (zero time-lag, absolute value), SUSHI and super-localization reconstructions, respectively. Red and cyan colored blood vessels correspond to negative and positive flow with respect to the transducer. Judging visually, the SOFI image seems clearer than the temporal mean image, though the resolution enhancement is limited. The SUSHI reconstruction is very well-defined and sharp, depicting clear bifurcations and intertwined blood vessels. In contrast, the super-localization image seems noisier without a clear depiction of the bifurcation. This happens since in this scan UCA overlap is present, which emphasizes its limitation in the case where high densities are used to reduce acquisition time bellow 100ms.

Panels (e), (f) present selected intensity profiles of the four methods along the solid and dashed yellow lines, respectively (profiles were taken with respect to the red colored blood vessels). These panels support the conclusions drawn from the visual comparison. Panel (e) shows that SUSHI (right arrow, green) clearly separates the two leftmost blood vessels. The temporal mean and SOFI profiles (plus head, black and diamond blue, respectively) do not exhibit such separation. The super-localization profile does not produce a clear depiction of the bifurcation. On the other hand, Panel (f) demonstrates that when isolated blood vessels are considered and clear UCA separation is evident, the resolution of SUSHI, even on experimental scans, is similar to that of the super-localization recovery, though slightly lower.

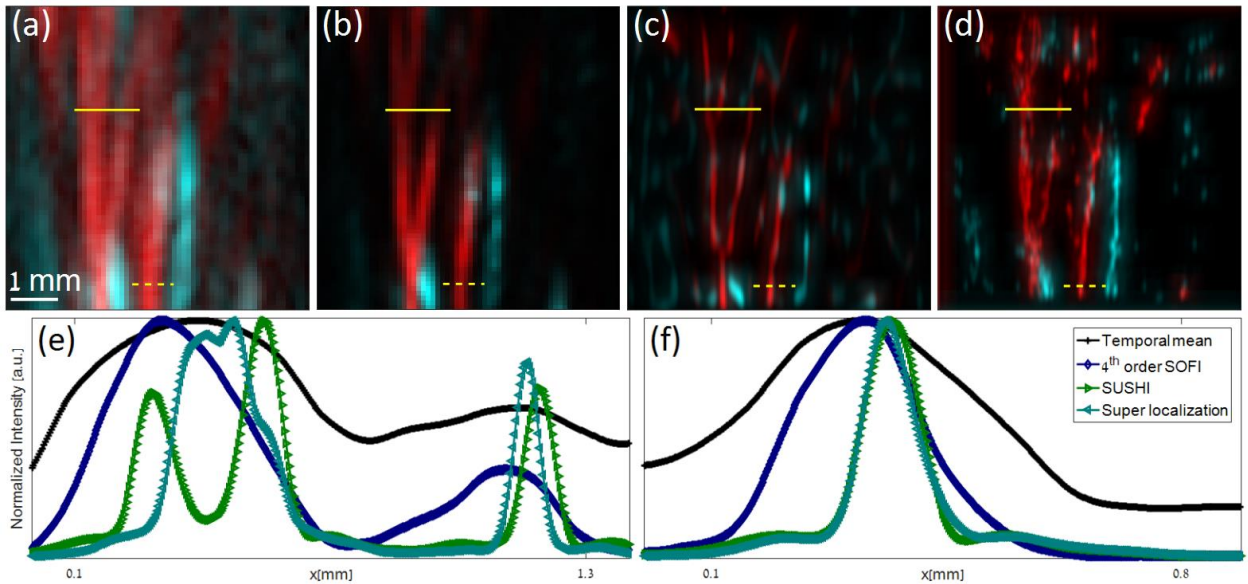

Fig. 3: **Spatial resolution comparison.** Blood vessel bifurcations, appearing in a scan acquired with high concentration of contrast agents, are depicted as (a) temporal mean, using 150 frames, (b) 4th order SOFI, (c) SUSHI and (d) super-localization. Red and cyan colors depict hemodynamic flow from and towards the transducer, respectively. Comparing visually, the super-localization image seems very noisy and unclear, while temporal mean and SOFI images have degraded resolution compared with SUSHI. SUSHI exhibits the clearest image of the vasculature, presenting a sharp, almost noiseless reconstruction, compared with the other methods. (e)-(f) intensity profiles measured along the solid and dashed yellow lines on panels (a)-(d). All profiles were taken with respect to the red blood vessels only. Panel (e) shows that in high density areas (e.g. bifurcations), SUSHI is superior, while in low density areas (e.g. isolated vessel in panel f), SUSHI exhibits comparable spatial resolution to super-localization.

The measured ratios between the temporal mean FWHM and the SUSHI/super-localization FWHM is 4.55 and 4.97, respectively, almost 5 times better than that of the original temporal mean. Here, the change in FWHM does not reflect the full increase in resolution since it is affected by the width of the blood vessels.

After comparing the spatial resolution of SUSHI and super-localization images, we proceed to demonstrate the ability of SUSHI to produce sub-diffraction movies with a high temporal resolution of $25Hz$, capturing changes in the imaged flow pattern. Figure 4, panels (a)-(d) present four 4th order SOFI images (zero time-lag) from a longer rabbit scan, injected with the contrast agent concentration similar to that used in the clinic. Each panel is composed from 100 consecutive frames from the 1000 frames included in the complete scan. Corresponding SUSHI reconstructions are presented in panels (e)-(h). The four high resolution SUSHI images illustrate a temporal resolution of $25Hz$. The white arrows in panels (f) and (g) point to a clear bifurcation which is considerably less visible, and with poorer resolution in the corresponding SOFI images. This bifurcation vanishes in panel (h), suggesting a difference in the vasculature captured during later parts of this sub-second scan. This observation shows the ability of SUSHI to monitor hemodynamic changes in a high spatio-temporal resolution.

Panel (i) illustrates the spatial shift (~0.5mm) in the position of the blood vessel (panel (e) in green and panel (h) in purple), indicated by the white arrow, over a span of $240ms$ ($40ms$ frame-rate). Since super-localization based techniques operate with longer acquisition periods, they require motion compensation to be applied to the localizations to reduce the overall localization error. SUSHI does not require such compensation in order to provide super-resolved time-lapse imaging of fast hemodynamic changes, but might benefit from it if the acquisition is extended when aiming to maximize the percentage of the vasculature imaged during the scan of low perfused tissues, such as tumors.

## VII. DISCUSSION

In this work, SUSHI is shown to produce images with exceptional combination of spatial super-resolution and high temporal resolution. This is achieved by relying on the ability to separate close-by vessels according to their Doppler velocities; the independence of CEUS fluctuations originating from different vessels; and by exploiting sparsity in the correlation domain. This prior information enables the proposed framework to produce an unprecedented spatiotemporal resolution trade-off: a 10-fold increase in spatial resolution and high temporal resolution. In cases where high UCA concentrations are considered, SUSHI produces a clear depiction of the vasculature with sub-second acquisition times, compared to typically a few minutes in super-localization scans [6], [8]. When well isolated blood vessels were analyzed, FWHM values of SUSHI were comparable to those of super-localization although slightly lower.

Fast super-resolution opens a vast range of opportunities for future applications and follow-up studies. Foremost, it facilitates functional (hemodynamic) super-resolution imaging that could bridge, for example, between cerebral anatomical super-resolution and functional neural imaging. In addition, it solves several important practical limitations that currently hinder wide clinical use of current CEUS super-resolution techniques. First, SUSHI acquisitions can be performed between breaths without the need for long breath-holds, which are unmanageable for many patients. Second, the numerically efficient implementation described in Appendix A2, and the potential for parallel processing of many image patches, could facilitate an in-clinic fast reconstruction process. Finally, when using scanners with limited computational resources, a reduction of almost 2 orders in the number of acquired frames compared to super-localization techniques means lower amounts of data to store and process.

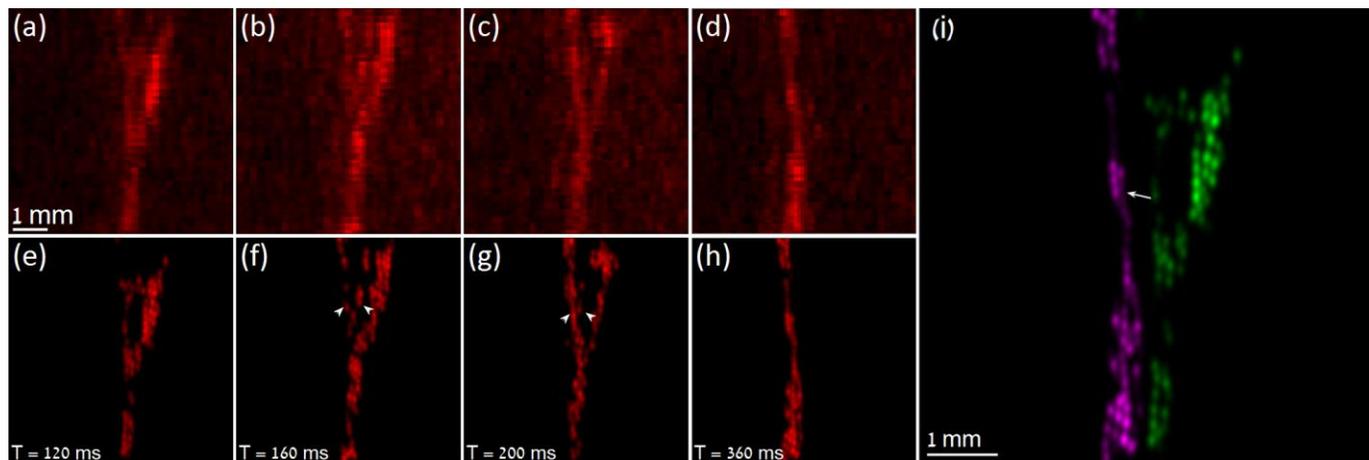

Fig. 4: **Demonstration of the high temporal resolution of SUSHI**. Panels (a)-(d) illustrate four 4th order SOFI snapshots (zero time-lag) of the same bifurcation of a blood vessel in a kidney of a white New-Zealand rabbit, injected with ultrasound microbubbles, at different times. Here, very short ensembles of 100 frames were used for each image. The times of the snapshots are given in the lower left corners of panels (e)-(h), which present the corresponding SUSHI reconstructions to panels (a)-(d). Panel (i) shows the super-imposed image of panels (e) and (h) in green and purple, respectively, spanning a temporal duration of 240ms (40ms frame-rate). Bifurcation of blood vessels marked by the white arrow heads can be clearly seen in panels (f) and (g), which is clearly less visible in the corresponding SOFI panels, (b) and (c). The white arrow in panel (i) serves to emphasize the lateral translation of the blood vessel during a period of 400ms. This example depicts the ability of SUSHI to image in vivo fine vasculature with a high spatio-temporal resolution.

SUSHI shares an inherent physical limitation of all CEUS imaging methods with short acquisition times, even when high ultrasound contrast agent concentrations are used: certain vessels with very low flow velocities might not contain microbubbles during the imaging interval. This is also true for both SUSHI and CEUS imaging methods with lower CEUS concentrations and longer acquisition times, no matter what processing method is applied. Although the use of high concentrations of ultrasound contrast agents in this work maximizes the portion of the vessels included in the scan within a given acquisition time, it cannot ensure a full coverage of the vasculature if very short acquisition intervals are needed. Incorporating microbubbles tracking over several frames within the super-localization framework, can improve microbubble detection and ULM estimations which could be important specifically in short CEUS acquisitions [48]. The integration of sparsity and UCA motion kinematics is a topic of continuing research [50].

Finally, in this study SUSHI was implemented using fast plane-wave imaging to detect fast hemodynamic changes. Currently, the use of fast plane-wave imaging is not widespread, mainly due to hardware limitations. When considering variance estimation, as was presented here, SUSHI can in principle also be applied with lower-rate commercially available scanners, using wide beam imaging for example. In the general sense, sparsity can even be used in the processing of images when RF signals are not accessible [50].

## VIII. CONCLUSION

This work presents a new and improved tradeoff between spatial and temporal resolutions in CEUS imaging: short acquisitions of only tens of milliseconds, with a $25Hz$ temporal resolution, and improved super-resolution abilities with 64-fold increase in pixel density and up to a 10-fold increase in spatial resolution. Drawing inspiration from SPARCOM, these results are achieved by a compressed sensing framework that combines prior knowledge on the temporal fluctuations of the received signal, alongside the sparse nature of the underlying signal. The proposed approach is characterized by short acquisition times, computationally efficient implementation, and reduced memory burden, which together could simplify the clinical adaptation of super-resolution CEUS. In addition, the enhanced spatiotemporal resolution provides researchers with a new set of tools that may enable, for example, the investigation of malignant hemodynamic patterns, super-resolution imaging of cardiac vasculature, and the monitoring of fast hemodynamic changes in functional neural scans.

## APPENDIXES

### A1. SUPPORTING FIGURES FOR THE THEORETICAL DERIVATIONS

Figure 5 illustrates spectral decomposition of the continuous Doppler spectrum into several bands for several possible selections of the temporal bandpass filters, described in (10). The upper illustration depicts two symmetric filters $D_1$ and $D_2$, separating the spectrum into positive flow and negative flow, with respect to the transducer. The lower figure shows four filters, $D_1 - D_4$, separating the spectrum into four categories. This decomposition provides both flow direction and in each direction two flow categories. For example, $D_3$ indicates slow, positive flow with respect to the transducer, while $D_4$ indicates fast, positive flow. The same holds for $D_2$ and $D_1$ for negative flow. Each such filter is applied on every pixel's time-trace, separately. Additional decompositions can also be performed, such that each filtered signal occupies a narrower velocity range.

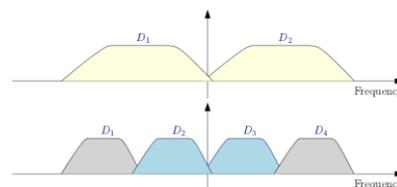

Fig. 5: **Separation to Doppler frequency bands using temporal bandpass filters.** Upper panel: Two Doppler filters. Lower panel: Four Doppler filters.

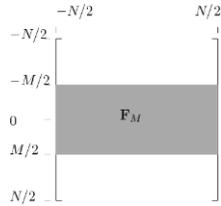

Fig. 6: **Partial discrete Fourier matrix of low frequency components.**

Fig. 6 depicts the partial discrete Fourier matrix, as described in Section IV-A. The matrix $\boldsymbol{F}_M$ is constructed by first generating a full $N{\times}N$ DFT matrix and then removing some of its rows. In this illustration, the zero frequency is at the middle of the matrix, negative frequencies correspond to the upper half of rows, and positive frequencies correspond to the lower half of rows. The matrix $\boldsymbol{F}_M$ consists only of the $M$ rows from $-M/2$ to $M/2$, as marked by the gray rectangle.

## A2. SUPPORTING MATERIAL FOR IMPLEMENTATION OF SUSHI

### A. Use of higher statistical order in SUSHI processing

Sparse recovery can be used not only for the correlations image, but theoretically for any statistical image, e. g. 4th order images, in which the value of each pixel is the 4th moment of its time-trace. Thus, (14) in the main text will consist of a superposition of the absolute PSF $|h|$ raised to the power of 4 instead of 2, and $g_{i_x,i_z}$ will consist of the 4th order moment estimation of the emitters' fluctuations. SUSHI will then be applied using an estimate of the PSF raised to the power of 4 when minimizing (17)-(19).

In practice, we refrain from high-order statistical estimation and demonstrate SUSHI using correlations only, since statistical estimation of high-order moments requires an exponentially increasing number of frames to retain the same SNR level. This in turn reduces the temporal resolution of SUSHI, since longer movie ensembles are required for the estimation process. Thus, for the sake of simplicity and to achieve good temporal resolution, we restricted the demonstration in this paper to second order statistics, which can be estimated from a relatively low number of frames, as we demonstrate in Section VI of the main text. There we show a significant improvement in the spatial resolution compared to the diffraction limit, with a sub-second temporal resolution.

### B. Forward problem in matrix-vector form

For convenience of the reader we present here the forward problem (16) in matrix-vector form in a more explicit way. Specifying separately each element of the PSF matrix $\boldsymbol{H}$ and the vectors $\boldsymbol{y}$ and $\boldsymbol{x}$ we get

$$
\begin{bmatrix} y_1(\tau) \\ \vdots \\ y_M(\tau) \end{bmatrix} = \begin{bmatrix} H[0,0] & 0 & 0 \\ 0 & \ddots & 0 \\ 0 & 0 & H[M-1,M-1] \end{bmatrix} (\boldsymbol{F}_M \otimes \boldsymbol{F}_M) \begin{bmatrix} x_1(\tau) \\ x_2(\tau) \\ \vdots \\ x_N(\tau) \end{bmatrix}, \quad (20)
$$

where $\boldsymbol{y}$ is the temporal autocorrelation vector, calculated from the data measured on the low-resolution grid, and $\boldsymbol{x}$ is the sparse representation vector that we want to estimate (on the high-resolution grid). $M$ is the width and length of the low-resolution grid, $N$ is the width and length of the high-resolution grid and $\tau$ is the discrete time lag used for the autocorrelation calculation.

### C. PSF estimation

In this section, we provide an example of the estimated PSF, using the procedure described in Section IV-B. An example of the real component of an automatically detected patch with a recoverable microbubble is depicted in Fig. 7a. The modulation of the pulse can be clearly seen. The final estimated PSF is presented in Fig. 7b in its absolute value. The amplitude of cross sections through this PSF, in the lateral and axial directions, are presented in Fig. 7c and Fig. 7d, respectively, along with their Gaussian approximation. The limitation of the Gaussian approximation is evident by looking at the additional details appearing in the estimated PSF. These details include side lobes (Fig. 7c) in the lateral direction and asymmetric structure, related to the ringing of the transducer, in the axial direction (Fig. 7d). The estimated PSF is then used in the SUSHI algorithm to recover the underlying vasculature with sub-diffraction resolution.

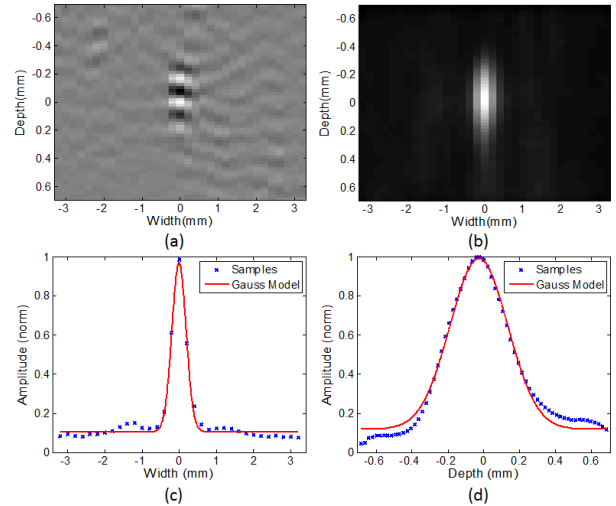

Fig. 7: **PSF estimation from resolvable microbubbles.** (a) A patch including a single resolvable microbubble (the real part of the signal is displayed). (b) The absolute value of the final estimation of the PSF. (c) Lateral section through the estimated PSF showing side-lobes. (d) Axial cross section through the estimated PSF showing the ringing of the transmitted pulse. Clearly, there is some mismatch to the simplified Gaussian model. However, SUSHI can operate adequately with both the estimated PSF (e.g. Fig. 3) and a Gaussian estimate (e.g. Fig. 4).

## A3. SUPPORTING IN-VIVO RESULTS

### A. Additional in-vivo example

In this section, we show an additional example of the high-resolution reconstruction obtained by SUSHI, when high densities are used. Figure 8 shows an additional patch of sub diffraction sized blood vessels, taken from the same kidney scans (150 frames) presented in Figs. 2, 3 in the main paper. The vessels in red contain negative flow with respect to the transducer, while vessels in cyan illustrate the positive flow.

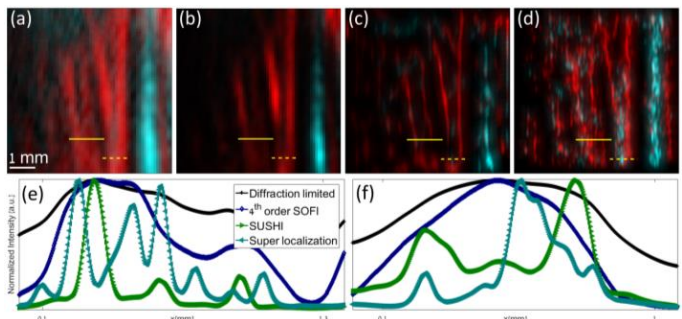
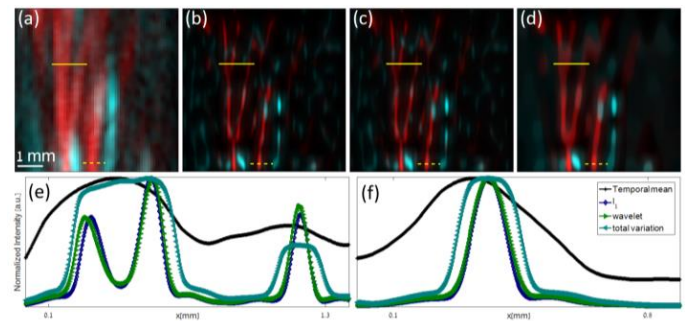

Fig. 8: **Additional spatial resolution comparison.** Blood vessel bifurcations are depicted with temporal mean, using 150 frames (a), 4th order SOFI (b), SUSHI (c) and super-localization (d). Red and cyan colors depict hemodynamic flow from and towards the transducer. Plots (e) and (f) show two intensity profiles measured along the solid and dashed yellow lines on panels (a)-(d). All profiles were taken with respect to the red blood vessels only.

Panels (a)-(d) depict four reconstructions: temporal mean, 4th order SOFI (zero time-lag, absolute value), SUSHI and super-localization, respectively. Judging visually, the SUSHI recovery seems the sharpest of all recoveries, showing clear bifurcations, which are missing in the temporal mean and SOFI recoveries. Since the microbubbles density is high (clinical dose), the super-localization technique involves many false detections and results in a noisy and unclear image. Panels (e), (f) depict selected intensity profiles along the solid and dashed yellow lines, respectively. The temporal mean profile is given by the plus-head, black line, 4th order SOFI by the blue diamond-head line, SUSHI by the green right-arrowhead and super-localization by the turquoise, left-arrowhead. Both panels serve to illustrate that SUSHI detects sub-diffraction blood vessels, which are not depicted by the temporal mean and SOFI reconstructions, while achieving clearer depiction compared to super-localization, when high concentrations are used. A clear depiction of bifurcations is seen in the SUSHI profiles (e.g. 3 blood vessels on panel (e)), while the super-localization profiles seem noisy and contain many false detections of blood vessel.

### B. Reconstruction with different sparsity priors

In this section, we provide a comparison between SUSHI recoveries under different sparsity assumptions. We compare with the same data presented in Fig. 3. In all cases, sparse recovery was performed using $\lambda = 0.5$ and 150 iterations. Figure 9 shows the temporal mean image in panel (a) and SUSHI recoveries in panels (b)-(d). Reconstructions were performed using (17), (19) and (18), respectively ($l_1$, wavelet and total-variation). The wavelet filter used in the reconstruction of panel (c) is a Daubechies wavelet with 16 taps, and we use a single level of decomposition. Clearly, panels (b) and (c) show similar images and achieve similar spatial resolution. The reconstruction in panel (d), using the total-variation norm yields a reconstruction with poorer spatial resolution. Panels (e) and (f) show intensity profiles which further support our analysis. Both recoveries in panels (b) and (c) ($l_1$ and wavelet) managed to resolve the left bifurcation in panel (e), while the reconstruction using the total-variation norm did not (though it achieved better resolution than the temporal mean image). Even in isolated blood vessels (e.g. panel (f)), the width of the $l_1$ and wavelet reconstructions was similar and narrower than the total-variation reconstruction.

Fig. 9: **Sparse recovery under different sparsity assumptions**. Blood vessel bifurcations are depicted with temporal mean, using 150 frames (a), SUSHI recovery using (17) (b), SUSHI recovery using (19) (c) and SUSHI recovery using (18) (d). Red and cyan colors depict hemodynamic flow from and towards the transducer. Intensity profiles are given in panels (e) and (f), corresponding to the solid and dashed yellow lines in panels (a)-(d), respectively. Comparing visually, panels (b) and (c) show similar reconstruction ($l_1$ and wavelet), while panel (d) (total-variation) shows a reconstruction with degraded resolution. Panels (e) and (f) support this observation. For example, panel (e) shows a clear bifurcation for the $l_1$ and wavelet recoveries, which is absent in the total-variation image.

### A4. ADDITIONAL SIMULATIONS

#### A. Increased microbubble densities in numerical simulations

Increasing the concentration of microbubbles injected into the body increases the probability of at least one microbubble flowing inside a given vessel during the duration of the ultrasound scan. Therefore, using the highest clinically approved concentration maximizes the portion of the vasculature detected during a certain imaging interval. This is true no matter what processing method is applied to the acquired CEUS data. Therefore, with increased temporal resolution being the focus of our study, the maximal clinically approved ultrasound contrast agent concentration is used in all the in vivo scans. To test the capabilities of this method over a wide range of concentrations, numerical simulations with a different number of microbubbles were produced.

In this section, we compare the performance of SUSHI

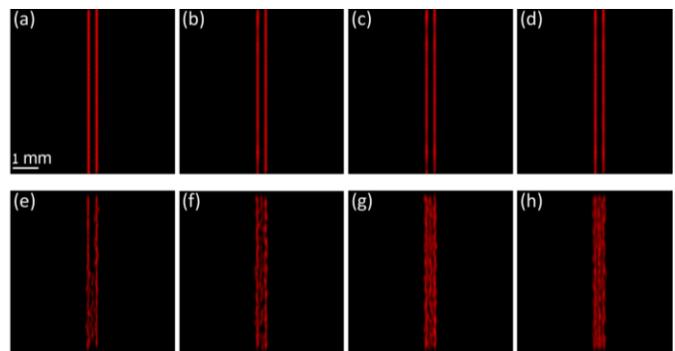

Fig. 10: **Simulation results of increased microbubble densities.** (a)-(d) correspond to SUSHI recoveries of the two streamlines presented in Fig. 1, and recoveries with increasing microbubble density of 2,3 and 4 times the original concentration, respectively. (e)-(h) Corresponding super-localization recoveries. It can be observed that as the density of flowing microbubbles increases, super-localization techniques fail due to the strong overlap between the echoes of the bubbles. On the other hand, in all cases, SUSHI manages to recover the two streamlines similarly, for every density presented here.

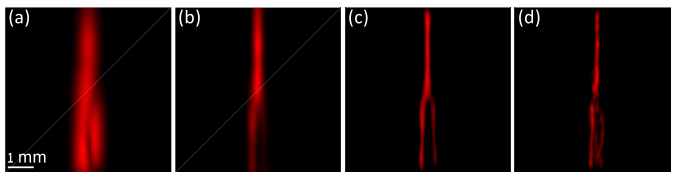

Fig. 11: **Simulation results of bifurcating blood vessel**. Blood vessel bifurcations are depicted with (a) a single frame from the movie, absolute value, (b) 4th order SOFI recovery, zero time-lag, (c) SUSHI recovery using Eq. (17), and (d) super-localization recovery. SUSHI manages to clearly detect the bifurcation better than the SOFI image, while showing its clear depiction, unlike the super-localization image.

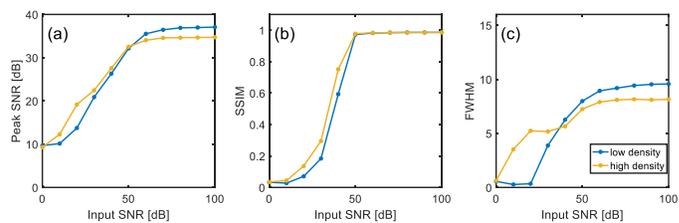

Fig. 12: **PSNR, SSIM and FWHM values for low and high UCA density simulations.** Panel (a) shows PSNR values in dB for SUSHI, while panel (b) illustrates SSIM values. Panel (c) presents FWHM ratios for the SUSHI reconstructions. It can be observed that, as expected, as the SNR increases, the metrics values for SUSHI increase and achieve high PSNR and SSIM values, as well as high values of FWHM ratios, for both concentrations.

against super-localization using data generated by a set of simulations with increased microbubble densities, moving along the same two streamlines presented in Fig. 1. Four movies were generated, 150 frames each, of microbubbles flowing along the two streamlines. In the original simulation, three Gaussian bubbles were positioned in the right vessel and six in the left vessel. Here, each movie has increased microbubbles density. The second movie has twice the density compared to the first, the third movie has three times the density and the fourth movie has four times the density. Panels (a)-(d) in Fig. 10 show the corresponding SUSHI reconstructions, while panels (e)-(h) illustrate the corresponding super-localization recoveries, for increasing densities, respectively. All reconstructions were smoothed with the same Gaussian kernel, so that the comparison is performed under similar conditions.

Judging the recoveries in panels (a)-(h), one observes that for all these concentrations SUSHI produces successful reconstructions, while a clear degradation in the super-localization estimations is evident. In panels (g) and (h), one can note many false-positives, and it is nearly impossible to clearly separate the two streamlines, thus demonstrating the limitations of super-localization when applied to CEUS scans with high CEUS concentrations. Figures. 8 and 10 substantiate the conclusions drawn that when using high densities, the performance of SUSHI is superior compared to the other methods. Higher concentrations lead to a reduction in the number of frames required to create a single super-resolved image, which in turn leads to an increase in the temporal resolution.

### B. Bifurcation simulation

In this section, we provide an additional simulation comparison between SUSHI, SOFI and super-localization, of a bifurcation blood vessel. Similar to the first bifurcation in [48], the width of the main vessel is 40μm with a peak velocity of 15mm/s and the width of the secondary vessels was 25 μm. The PSF was identical to the one in Fig. 1 and Fig. 10, and a total of 600 frames were simulated. Initial location along the cross section of the vessel was uniformly randomly selected. Temporal density is identical to the simulation of two parallel blood vessels (Fig. 1) and in the branches of the bifurcating vessels, on average. Panels a-d of Fig. 11 show a single frame from the movie, where significant UCA overlap is evident, 4th order SOFI image, SUSHI image and super-localization recovery. Similar conclusions can be deduced here also. Clearly SUSHI outperforms the SOFI and temporal mean images, by clearly detecting the bifurcation (much earlier than the SOFI image). The SUSHI recovery is also better than the super-localization recovery, in which the depiction of the bifurcation in unclear and with artifacts.

### A5. Noise analysis

We next use two of the simulations presented in Fig. 1 (lowest and highest UCA concentration simulations) to test the performance of SUSHI with varying SNR conditions. To perform a fair analysis, we normalize the intensity of each movie frame to 1 and define the SNR as $SNR = 1/\sigma^2$, where σ is the standard deviation of the noise. We add white Gaussian noise for increasing SNR values to the movie and performed SUSHI recovery. Each recovery is then smoothed with the same kernel as before (post-processing) and its peak intensity is normalized to one. We compute three metrics, peak SNR (PSNR) and structural similarity index (SSIM) against the noiseless SUSHI recovery, and the full width at half maximum (FWHM) ratio computed as the ratio between the FWHM value of the temporal mean image, divided by the FWHM of the corresponding SUSHI recovery.

Figure 12 depicts these values as a function of increasing SNR. It can be observed from all panels that as the SNR increases, the quality of the SUSHI recoveries, as measured by the PSNR, SSIM and FWHM ratio metrics increases, for both concentrations. For SNR values of 20-30dB and higher, SUSHI performs well and achieves PSNR values above 20dB, SSIM values close to one (not really affected by the different concentrations) and a FWHM ratio of ~8-10, indicating a resolution gain of ~8-10 times better than the temporal mean image, for both concentrations. As the concentration increases, the FWHM ratio decreases (Panel (c)), but still, an order of magnitude improvement in the FWHM ratio is evident for high SNR values.

### References

[1] J. M. Hudson *et al.*, "Dynamic contrast enhanced ultrasound for therapy monitoring," *Eur. J. Radiol.*, vol. 84, no. 9, pp. 1650–1657, 2015.

[2] C. B. Conti, M. Giunta, D. Gridavilla, D. Conte, and M. Fraquelli, "Role of Bowel Ultrasound in the Diagnosis and Follow-up of Patients with Crohn's Disease," *Ultrasound Med. Biol.*, vol. 43, no. 4, pp. 725–734, 2017.

[3] C. N. Hall *et al.*, "Capillary pericytes regulate cerebral blood flow in health and disease.," *Nature*, vol. 508, no. 7494, pp. 55–60, 2014.

[4] E. Mace E., G. Montaldo, I. Cohen, M. Baulac, M. Fink, and M. Tanter, "Functional ultrasound imaging of the brain.," *Nat Methods*, vol. 8, no. 8, pp. 662–664, 2011.

[5] R. K. Jain, "Normalization of tumor vasculature: an emerging concept in antiangiogenic therapy.," *Science (80).*, vol. 307, no. 5706, pp. 58–62, 2005.

[6] C. Errico *et al.*, "Ultrafast ultrasound localization microscopy for deep super-resolution vascular imaging," *Nature*, vol. 527, no. 7579, pp. 499–502, 2015.

[7] M. a. O'Reilly and K. Hynynen, "A super-resolution ultrasound method for brain vascular mapping.," *Med. Phys.*, vol. 40, no. 11, p. 110701, 2013.

[8] K. Christensen-Jeffries, R. J. Browning, M. X. Tang, C. Dunsby, and R. J. Eckersley, "In vivo acoustic super-resolution and super-resolved velocity

mapping using microbubbles," *IEEE Trans. Med. Imaging*, vol. 34, no. 2, pp. 433–440, 2015.

[9]    O. Couture, B. Besson, G. Montaldo, M. Fink, and M. Tanter, "Microbubble ultrasound super-localization imaging (MUSLI)," *IEEE Int. Ultrason. Symp. IUS*, no. Id 797, pp. 1285–1287, 2011.

[10]   F. Lin, S. E. Shelton, D. Espíndola, J. D. Rojas, G. Pinton, and P. A. Dayton, "3-D ultrasound localization microscopy for identifying microvascular morphology features of tumor angiogenesis at a resolution beyond the diffraction limit of conventional ultrasound," *Theranostics*, vol. 7, no. 1, pp. 196–204, 2017.

[11]   J. Foiret, H. Zhang, T. Ilovitsh, L. Mahakian, S. Tam, and K. W. Ferrara, "Ultrasound localization microscopy to image and assess microvasculature in a rat kidney," *Sci. Rep.*, vol. 7, no. 1, pp. 1–12, 2017.

[12]   M. Siepmann, G. Schmitz, J. Bzyl, M. Palmowski, and F. Kiessling, "Imaging tumor vascularity by tracing single microbubbles," *IEEE Int. Ultrason. Symp. IUS*, pp. 1906–1908, 2011.

[13]   E. Betzig *et al.*, "Imaging intracellular fluorescent proteins at nanometer resolution.," *Science*, vol. 313, no. 5793, pp. 1642–5, 2006.

[14]   M. J. Rust, M. Bates, and X. W. Zhuang, "Sub-diffraction-limit imaging by stochastic optical reconstruction microscopy (STORM)," *Nat Methods*, vol. 3, no. 10, pp. 793–795, 2006.

[15]   A. Urban, C. Dussaux, G. Martel, C. Brunner, E. Mace, and G. Montaldo, "Real-time imaging of brain activity in freely moving rats using functional ultrasound.," *Nat. Methods*, vol. 12, no. july, pp. 873–878, 2015.

[16]   S. B. Gay, C. L. Sistrom, C. A. Holder, and P. M. Suratt, "Breath-Holding Capability of Adults: Implications for Spiral Computed Tomography, Fast-Acquisition Magnetic Resonance Imaging, and Angiography.," *Invest. Radiol.*, vol. 29, no. 9, pp. 848–851, 1994.

[17]   S. R. Wilson, H. J. Jang, K. K. Tae, H. Iijima, N. Kamiyama, and P. N. Burns, "Real-time temporal maximum-intensity-projection imaging of hepatic lesions with contrast-enhanced sonography," *Am. J. Roentgenol.*, vol. 190, no. 3, pp. 691–695, 2008.

[18]   T. Dertinger, R. Colyer, G. Iyer, S. Weiss, and J. Enderlein, "Fast, background-free, 3D super-resolution optical fluctuation imaging (SOFI).," *Proc. Natl. Acad. Sci. U. S. A.*, vol. 106, no. 52, pp. 22287–92, 2009.

[19]   A. Bar-Zion, C. Tremblay-Darveau, O. Solomon, D. Adam, and Y. C. Eldar, "Fast Vascular Ultrasound Imaging with Enhanced Spatial Resolution and Background Rejection," *IEEE Trans. Med. Imaging*, vol. 7, pp. 1–12, 2016.

[20]   A. Bar-Zion, O. Solomon, C. Tremblay-Darveau, D. Adam, and Y. C. Eldar, "Sparsity-based ultrasound super-resolution imaging," in *Proceedings of the 23rd European symposium on Ultrasound Contrast Imaging*, 2017, pp. 156–157.

[21]   A. Beck and M. Teboulle, "A Fast Iterative Shrinkage-Thresholding Algorithm," *SIAM J. Imaging Sci.*, vol. 2, no. 1, pp. 183–202, 2009.

[22]   Z. Tan, Y. C. Eldar, A. Beck, and A. Nehorai, "Smoothing and decomposition for analysis sparse recovery," *IEEE Trans. Signal Process.*, vol. 62, no. 7, pp. 1762–1774, 2014.

[23]   D. L. Donoho, "Compressed sensing," *IEEE Trans. Inf. Theory*, vol. 52, no. 4, pp. 1289–1306, 2006.

[24]   Y. C. Eldar and G. Kutyniok, *Compressed sensing: theory and applications*. Cambridge University Press, 2012.

[25]   O. Solomon, M. Mutzafi, M. Segev, and Y. C. Eldar, "Sparsity-based super-resolution microscopy from correlation information," *Opt. Express*, vol. 26, no. 14, pp. 18238–18269, 2018.

[26]   O. Solomon, Y. C. Eldar, M. Mutzafi, and M. Segev, "SPARCOM: Sparsity Based Super-Resolution Correlation Microscopy," arXiv preprint arXiv:1707.09255, 2017.

[27]   Y. C. Eldar, *Sampling Theory: Beyond Bandlimited Systems*. Cambridge University Press, 2015.

[28]   M. Elad, *Sparse and redundant representations: From theory to applications in signal and image processing*. 2010.

[29]   R. Baraniuk and P. Steeghs, "Compressive radar imaging," *IEEE Natl. Radar Conf. - Proc.*, pp. 128–133, 2007.

[30]   M. Lustig, D. Donoho, and J. M. Pauly, "Sparse MRI: The application of compressed sensing for rapid MR imaging," *Magn. Reson. Med.*, vol. 58, no. 6, pp. 1182–1195, 2007.

[31]   T. Chernyakova and Y. C. Eldar, "Fourier-domain beamforming: The path to compressed ultrasound imaging," *IEEE Trans. Ultrason. Ferroelectr. Freq. Control*, vol. 61, no. 8, pp. 1252–1267, 2014.

[32]   Y. Shechtman *et al.*, "Sparsity-based single-shot sub-wavelength coherent diffractive imaging," *2012 IEEE 27th Conv. Electr. Electron. Eng. Isr. IEEEI 2012*, vol. 11, no. 5, pp. 455–459, 2012.

[33]   P. Pal and P. P. Vaidyanathan, "Pushing the limits of sparse support recovery using correlation information," *IEEE Trans. Signal Process.*, vol. 63, no. 3, pp. 711–726, 2015.

[34]   D. Cohen and Y. C. Eldar, "Sub-Nyquist sampling for power spectrum sensing in Cognitive Radios: A unified approach," *IEEE Trans. Signal Process.*, vol. 62, no. 15, pp. 3897–3910, 2014.

[35]   C. Demené *et al.*, "Spatiotemporal Clutter Filtering of Ultrafast Ultrasound Data Highly Increases Doppler and fUltrasound Sensitivity," *IEEE Trans. Med. Imaging*, vol. 34, no. 11, pp. 2271–2285, 2015.

[36]   R. J. Eckersley, C. T. Chin, and P. N. Burns, "Optimising phase and amplitude modulation schemes for imaging microbubble contrast agents at low acoustic power," *Ultrasound Med. Biol.*, vol. 31, no. 2, pp. 213–219, 2005.

[37]   S. Bjærum, H. Torp, and K. Kristoffersen, "Clutter filter design for ultrasound

[38]   colour flow imaging," *IEEE Trans. Ultrason. Ferroelec., Freq. Contr.*, vol. 49, no. 2, pp. 204–209, 2002.

[38]   N. de Jong, *Acoustic Properties*. 2013.

[39]   D. H. Simpson, C. T. Chin, and P. N. Burns, "Pulse inversion Doppler: a new method for detecting nonlinear echoes from nonlinear microbubble contrast agents," *IEEE Trans. Ultrason. Ferroelectr. Freq. Control*, vol. 46, no. 2, pp. 372–382, Mar. 1999.

[40]   J. A. Jensen, "Estimation of Blood Velocities Using Ultrasound, A Signal Processing Approach." 1996.

[41]   R. Cohen and Y. C. Eldar, "Sparse Doppler Sensing Based on Nested Arrays," *IEEE Trans. Ultrason. Ferroelectr. Freq. Control, accepted*.

[42]   A. Bar-Zion, C. Tremblay-Darveau, M. Yin, and F. S. Foster, "Denoising of contrast enhanced ultyrasound cine sequences based on a multiplicative model," *IEEE Trans Biomed Eng*, vol. 62, no. 8, pp. 1969–1980, 2015.

[43]   C. Tremblay-Darveau *et al.*, "Improved contrast-enhanced Power Doppler using a coherence-based estimator," *IEEE Trans. Med. Imaging*, vol. 36, no. 9, pp. 1–1, 2017.

[44]   L. I. Rudin, S. Osher, and E. Fatemi, "Nonlinear total variation based noise removal algorithms," *Phys. D Nonlinear Phenom.*, vol. 60, no. 1, pp. 259–268, 1992.

[45]   A. Beck and M. Teboulle, "Fast gradient-based algorithms for constrained total variation image denoising and deblurring problems," *IEEE Trans. Image Process.*, vol. 18, no. 11, pp. 2419–2434, 2009.

[46]   C. a Schneider, W. S. Rasband, and K. W. Eliceiri, "NIH Image to ImageJ: 25 years of image analysis," *Nat. Methods*, vol. 9, no. 7, pp. 671–675, 2012.

[47]   M. Ovesný, P. Křížek, J. Borkovec, Z. Švindrych, and G. M. Hagen, "ThunderSTORM: A comprehensive ImageJ plug-in for PALM and STORM data analysis and super-resolution imaging," *Bioinformatics*, vol. 30, no. 16, pp. 2389–2390, 2014.

[48]   D. Ackermann and G. Schmitz, "Detection and tracking of multiple microbubbles in ultrasound B-mode images," *IEEE Trans. Ultrason. Ferroelectr. Freq. Control*, vol. 63, no. 1, pp. 72–82, 2016.

[49]   C. Tremblay-Darveau, R. Williams, L. Milot, M. Bruce, and P. N. Burns, "Combined perfusion and doppler imaging using plane-wave nonlinear detection and microbubble contrast agents," *IEEE Trans. Ultrason. Ferroelectr. Freq. Control*, vol. 61, no. 12, pp. 1988–2000, Dec. 2014.

[50]   O. Solomon, R. J. G. van Sloun, M. Wijkstra, M. Mischi, and Y. C. Eldar, "Exploiting flow dynamics for super-resolution in contrast-enhanced ultrasound," pp. 1–15, 2018.